PLOS one

# Finding Complex Biological Relationships in Recent PubMed Articles Using Bio-LDA

Huijun Wang[1], Ying Ding[2], Jie Tang[3], Xiao Dong[1], Bing He[2], Judy Qiu[4], David J. Wild[1]*

1 School of Informatics and Computing, Indiana University, Bloomington, Indiana, United States of America, 2 School of Library and Information Science, Indiana University, Bloomington, Indiana, United States of America, 3 Department of Computer Science, Tsinghua University, Beijing, China, 4 Pervasive Technology Institute, Indiana University, Bloomington, Indiana, United States of America

## Abstract

The overwhelming amount of available scholarly literature in the life sciences poses significant challenges to scientists wishing to keep up with important developments related to their research, but also provides a useful resource for the discovery of recent information concerning genes, diseases, compounds and the interactions between them. In this paper, we describe an algorithm called Bio-LDA that uses extracted biological terminology to automatically identify latent topics, and provides a variety of measures to uncover putative relations among topics and bio-terms. Relationships identified using those approaches are combined with existing data in life science datasets to provide additional insight. Three case studies demonstrate the utility of the Bio-LDA model, including association predication, association search and connectivity map generation. This combined approach offers new opportunities for knowledge discovery in many areas of biology including target identification, lead hopping and drug repurposing.

Citation: Wang H, Ding Y, Tang J, Dong X, He B, et al. (2011) Finding Complex Biological Relationships in Recent PubMed Articles Using Bio-LDA. PLoS ONE 6(3): e17243. doi:10.1371/journal.pone.0017243

Editor: Jörg Langowski, German Cancer Research Center, Germany

Received November 12, 2010; Accepted January 24, 2011; Published March 23, 2011

Copyright: © 2011 Wang et al. This is an open-access article distributed under the terms of the Creative Commons Attribution License, which permits unrestricted use, distribution, and reproduction in any medium, provided the original author and source are credited.

Funding: The authors have no support or funding to report.

Competing Interests: The authors have declared that no competing interests exist.

* E-mail: djwild@indiana.edu

## Introduction

Translational research in medicine is concerned with transforming basic laboratory science into effective patient therapies as quickly as possible. Developing effective treatments requires a cross-discipline understanding of medicine, pharmacology, biology and chemistry at the physiological, cellular, and molecular levels. Probably the most significant source of knowledge lies as text in published literature. PubMed [1] is an online resource that provides more than 19 million citations for published articles in journals and books, and while most are associated with short abstracts, an increasing number are now being accompanied by free, full text articles. At the same time, sophisticated interdisciplinary research has lead to the development and application of powerful methods to generate enormous amounts of new data resulting in an increased topical complexity of research articles. This complexity makes it challenging to efficiently discover, evaluate and synthesize the latest information, trends, and findings deposited in published literature in a reasonable amount of time. For the biomedical researcher, being able to quickly generate and ascertain the significance of associations between chemicals, genes and diseases are valuable in investigations relating to drug discovery. Thus, generating useful approaches to facilitate knowledge discovery through systematic analysis of abstracts and full-text journal articles is an important and ongoing challenge.

Natural language processing (NLP) is a common approach to text mining of biomedical corpora [2,3]. However, NLP largely relies on the syntactic and linguistic structure of documents, and is not in itself able to identify scientific relationships between terms. In contrast, statistical modeling techniques including Latent Dirichlet Allocation (LDA) [4] make the automated identification of topics from large document collections and corpora possible [5]. LDA, a hierarchical Bayesian model, has been extended to obtain relations between topics and terms [6]. Having a specialized and advanced LDA model using life sciences terms may provide a more effective way of exploring the biomedical literature.

Despite enormous investments in generating information pertinent to drug discovery and disease research, the problems associated with data integration are still a barrier to medical research [7]. An important tool in breaking down this barrier is the Semantic Web. Using Semantic Web technologies it becomes possible to convert data to a common syntax and specify the meaning of the data through shared vocabularies that can be specified as a formal, logical-based ontology. Bio2RDF [8], Chem2Bio2RDF[9], and Linked Open Drug Data (LODD) [10] are all projects involved in providing life sciences data using Semantic Web technologies. The resulting cloud of Linked Open Data now makes it possible to download interlinked data in a common format and the ability to query across their diverse resources, such resources are likely to become powerful drivers for increasing scientific productivity. Clearly, being able to link these datasets with complementary extracted information from the PubMed datasets would dramatically increase the overall opportunity for knowledge discovery.

The LDA model considered in this paper is a model for a text corpus viewed as a collection of bags of words. It assumes that people write an article with several topics in mind; each topic is associated with a different conditional distribution over a fixed set of words. A collection of documents can be seen as being generated by the same set of topics with different probability distribution for each document [4,11]. Therefore, LDA is a mixture model, *i.e.*, the mixture components are shared across all documents but each document exhibits different mixture proportions [4].

Since Blei et al. introduced the LDA model [4], various extended LDA models have been used in automatic topic extraction from text corpora. Rosen-Zvi et al. introduced the Author-Topic model which extended LDA to include authorship [5]. Each author is associated with a multinomial distribution over topics. They applied the model to a collection of 1,700 NIPS conference papers and 160,000 CiteSeer abstracts. The primary benefit of the author-topic model is that it allows the explicit inclusion of authors in the document models, providing a general framework for answering queries and making predictions at the level of authors as well as the level of documents. Based on Author-Topic model, McCallum and Wang presented an Author-Recipient-Topic (ART) model for social network analysis, which learns topic distributions based on the direction-sensitive messages sent between entities, adding the key attribute that distribution over topics is conditioned distinctly on both the sender and recipient [12]. Tang et al. further extended the LDA and Author-Topic model to the Author-Conference-Topic model [6], which is considered as a unified topic model to simultaneously model the different types of information in the academic network. They found that the proposed method had a high performance in expertise search and association search. Xiance and Maosong proposed a tag-LDA model which extended LDA model by adding the tag variable and applied it to social tagging systems [13].

In addition to text data, adapted LDA models are also applied to visual data for topic mining. In order to refine tags associated with images, Xu et al. proposed a regularized LDA (rLDA) which facilitates the topic modeling by exploiting both the statistical of tags and visual affinities of images in the corpus [14]. Wang and Grimson proposed a topic model Spatial Latent Dirichlet Allocation (SLDA), in which the knowledge of spatial structure can be flexibly added as a prior, grouping visual words which are close in space into the same document [15]. They found that SLDA achieved better performance than LDA when applied to a collection of images.

Those studies above extended the classic LDA model mainly by incorporating new variables to meet the customized demand in the applied area. Other advanced extensions of LDA model include supervised Latent Dirichlet Allocation (sLDA) [16] and dynamic topic model [12].

Some prior studies have been devoted to multiple alternatives of speeding up the learning of LDA, including parallelization across machines. Newman and et al. presented two synchronous methods, AD-LDA and HD-LDA, to perform distributed Gibbs sampling [17]. Asuncion, Smyth and Welling proposed asynchronous distributed learning algorithms for LDA and Hierarchical Dirichlet Process (HDP) in which processors independently perform Gibbs sampling on their local data and communicate their information in a local asynchronous manner with other processors [18]. Wang et al. introduced a parallel implementation of LDA on MPI and MapReduce, which smoothes out storage and computation bottlenecks and provides fault recovery for lengthy distribution computation [19].

As for applications of LDA in biomedical domain, Blei et al. examined 5,225 free-text items in the Caenorhabditis Genetic Center (CGC) Bibliography using the classic LDA model [20]. They found that like other graphical models for genetic, genomic and other types of biological data, the LDA model estimated from

CGC items had better predictive performance than two standard models (unigram and mixture of unigrams) trained using the same data. Zheng, et al. applied the classic LDA model to a corpus of protein-related MEDLINE titles and abstracts and extracted 300 major topics [21]. They found that those topics were semantically coherent and most represented biological objects or concepts. They further mapped those topics to controlled vocabulary of the Gene Ontology (GO) terms based on mutual information. They concluded that those identified topics provide parsimonious and semantically-enriched representation of the texts in a semantic space with reduced dimensionality that can be used to index text. Bundschus et al. presented a Topic-Concept model, which extends the basic LDA framework to reflect the generative process of indexing a PubMed abstract with terminological concepts from an ontology [22]. The Topic-Concept model extends the LDA framework by simultaneously modeling the generative process of document generation and the process of document indexing. For each of the concepts in the document a topic is uniformly drawn based on the topic assignments for each word in the document; each concept is sampled from a multinomial distribution over concepts specific to the sampled topic. They applied the model into a large-scale collection of medical text from PubMed and found that a number of important tasks for biomedical knowledge discover can be solved with Topic-Concept model.

While previous applications of LDA in the biomedical domain have yielded several benefits, few considered the extension of the LDA model to include bio-terms (that is a restricted vocabulary of genes, compounds, diseases, and so on) as input parameters. In this paper, we develop a Bio-LDA model, which extends the LDA model by incorporating bio-terms as input variables to the classic LDA model. The associations of the bio-terms are measured based on the topic distribution of the bio-terms. This approach is useful to establish hidden relations between biomedical concepts from literature compare to the commonly used co-occurrence-based methods [23,24]. The identified bio-term associations are evaluated using Chem2Bio2RDF.

Our contributions are: 1) the development of Bio-LDA, a novel advanced LDA model to mine the latent semantics among topics and biological terms; 2) conversion of identified latent semantics to RDF triples and alignment with existing semantic life data; 3) the demonstration of the application of these methods through use cases which cannot be solved by using traditional literature and database searches. This paper is organized as follows: Section 2 covers the data and methodology, the proposed Bio-LDA model, and other related tools/services built using this model. Section 3 presents the experimental results of the Bio-LDA model and describes three use cases to illustrate the utility of the approach in solving interesting problems in biomedical domain and section 4 offers our summary discussion.

## Materials and Methods

### Datasets

**Chem2Bio2Rdf.** Chem2Bio2RDF[9] consists of about 78 million RDF triples over 25 datasets relating to systems chemical biology, which is grouped into 6 domains, namely chemical (PubChem Compound, ChEBI, PDB Ligand), chemogenomics (KEGG Ligand, CTD Chemical, BindingDB, MATADOR, PubChem BioAssay, QSAR, TTD, DrugBank, ChEMBL, Binding MOAD, PDSP, PharmGKB), biological (UNIPROT, HGNC, PDB, GI), systems (KEGG Pathway, Reactome, PPI, DIP), phenotype (OMIM, Diseasome, SIDER, CTD diseases) and literature (MEDLINE/PubMed. Provenance information pertaining to these resources is available at http://chem2bio2rdf.org/datasets.html.

Chem2Bio2RDF data is linked to LODD and Bio2RDF data using owl:sameAs.

**MEDLINE and Bio-Terms Extraction.** PubMed offers a web-based and programmatic search service over its content [1]. However, this interface is limited to small- to medium-scale queries, and text mining using this interface is not possible. MEDLINE is the primary component of PubMed, where approximately 5400 biomedical journals published in the United States and worldwide, and covers abstracts from 1949 to present. The entire content of MEDLINE is available as a set of text files formatted in XML (eXtensible Markup Language). In this project, the 2010 MEDLINE/PubMed baseline database is used as our primary data source, which contains 617 files and 18,502,916 records (which covers citations through 2010).

In order to support our information extraction and text mining, we developed a system to load MEDLINE XML files to a relational database, extracting bio-terms from MEDLINE, and converting the relational database to RDF schema as shown in Figure 1.

The relational database schema used in our system is designed based on the category of bio-terms (compound, drug, gene, disease, side-effect, pathway) and DTD (Document Type Definition) provided by National Library of Medicine (NLM). Considering the size of MEDLINE database (over 18 million citations), we tried to minimize lookups by introducing redundant information in database. Our MEDLINE database contains several tables: *medline_citation* contains the title, abstract information, *medline_biblio* contains the bibliography information, *medline_author* contains the authors' information, *medline_comp* contains the mentioned compounds in the citation, *etc.* All tables contain a PubMed identifier (PMID) in one column, which connects tables and is also the key attribute in the RDF conversion.

Information such as citation, authors, journals, MeSH terms were directly parsed from the XML file and loaded into database. The bio-terms information needs to first apply extraction method to intermediate files and then load to database. In our system, we used the dictionary extraction method. Bio-term dictionaries are generated from the following data sources listed in Chem2Bio2RDF: the compound dictionary is generated from PubChem Synonym with the PubChem Compound identifier (CID); the drug dictionary is generated from DrugBank and used DBID as the identifier; the gene dictionary is generated from the HGNC and used UniprotID as the identifier; the disease dictionary is generated from the CTD (the comparative toxicogenomics database) and used MeshID as the identifier; the side effect dictionary is generated from the Sider and used UMLSID as the identifier; the pathway dictionary is generated from the KEGG pathway and used KeggID as the identifier. The extraction tools parses the XML file and extracts the terms based on the pre-generated dictionaries, then saved the results to intermediate flat files, and then loads the files into the database. Table 1 summarizes the dictionary attributes.

The D2R tool [25] is used to convert to the MEDLINE relational database to RDF schema and combined to the Chem2Bio2Rdf [9], which supports data visualization and complex query retrievals. The key attribute for the MEDLINE triples is the PMID. The extracted bio-terms, which used the well-

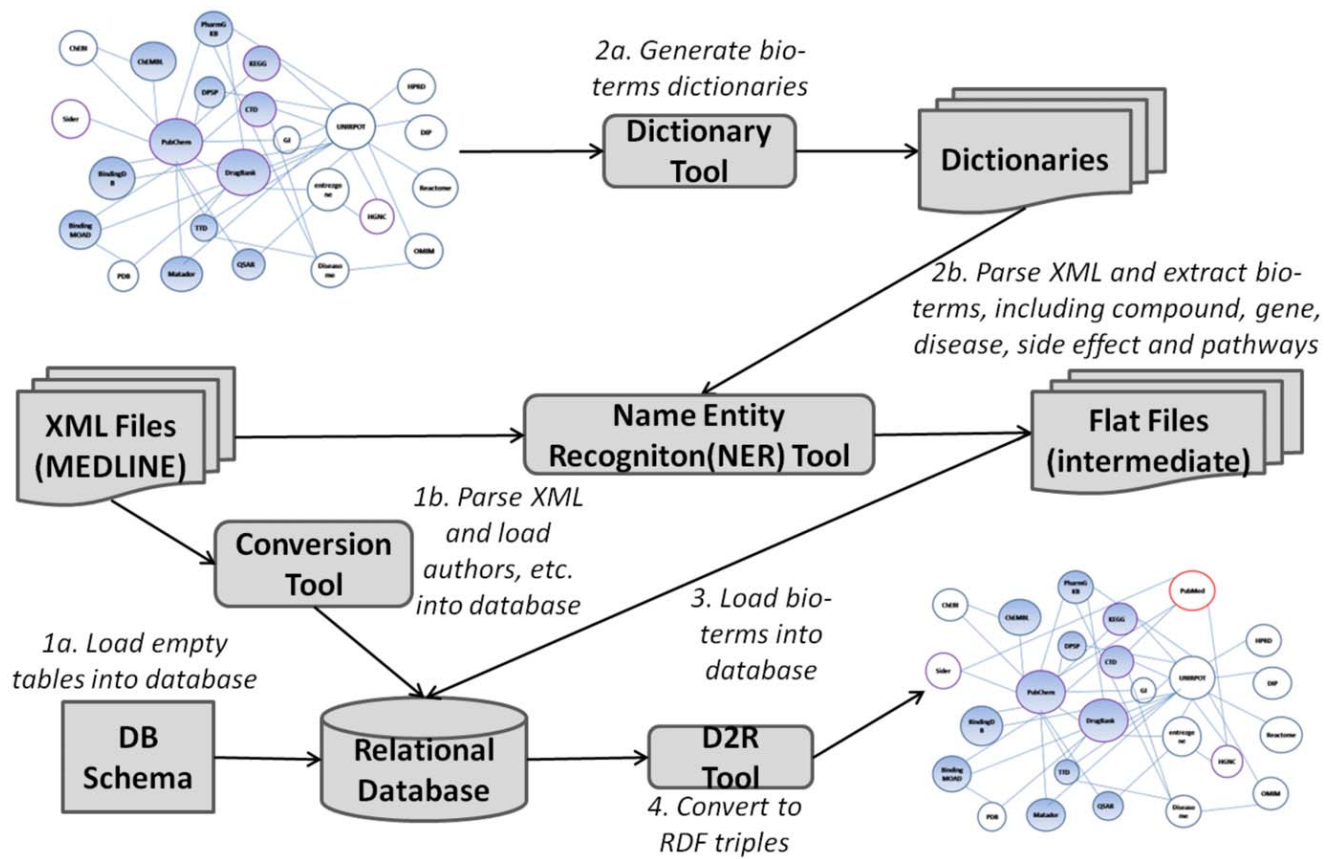

**Figure 1. Data Preprocessing.**
doi:10.1371/journal.pone.0017243.g001

**Table 1.** Statistics of the bio-terms extraction.

| Bio-Terms | # of unique terms | # of term-citation pairs | # of unique citations |
|---|---|---|---|
| Compound | 56,383 | 11,775,891 | 5,856,084 |
| Drug | 2,820 | 5,624,529 | 3,427,067 |
| Gene | 13,022 | 5,252,844 | 3,735,517 |
| Disease | 3,848 | 12,612,636 | 7,066,084 |
| Side Effect | 1,363 | 10,489,676 | 6,310,741 |
| Pathway | 180 | 916,754 | 838,090 |

doi:10.1371/journal.pone.0017243.t001

known identifies, are the key to bridge MEDLINE with other data sources, *i.e.* PubChem, UniProt, etc.

## Bio-LDA

The Bio-LDA model extends the ACT model [6] and emphasizes on bio-terms occurred in literatures. The basic assumption of the Bio-LDA model is that bio-terms of a paper would determine topics written in this paper and each topic then generates the words and determines the publication journal. The generative process can be summarized in the Figure 2:

1. For each bio-term $x = 1, \ldots, B$, draw $\theta_x \sim$ Dirichlet($\alpha$)
   For each topic $z = 1, \ldots, T$, draw $\phi_z \sim$ Dirichlet($\beta$), and $\psi_z \sim$ Dirichlet($\mu$)
2. For each document $d = 1, \ldots, D$
   Given the vector of bio-terms $\boldsymbol{b}_d$
   For each word $w_i$ in document $d$:

   Draw a bio-term $x_{di} \sim$ Uniform($\boldsymbol{b}_d$)
   Draw a topic $z_{di} \sim$ Dirichlet($\theta x_{di}$)
   Draw a word $w_{di} \sim$ Dirichlet($\phi z_{di}$)
   Draw a journal $j_{di} \sim$ Dirichlet($\psi z_{di}$)

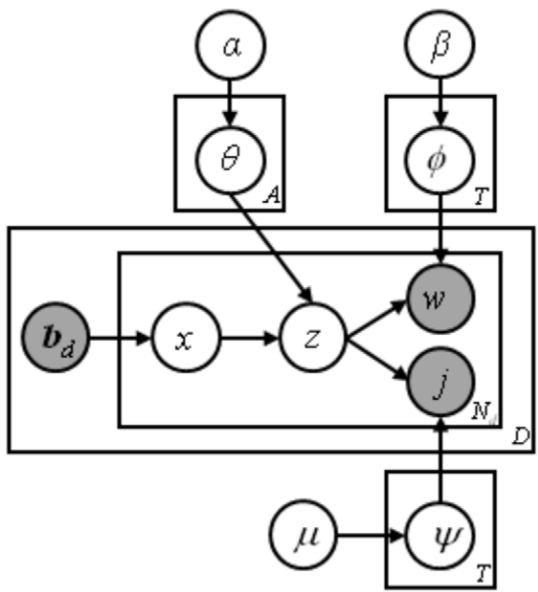

**Figure 2. Graphical representation of the Bio-LDA models.**
doi:10.1371/journal.pone.0017243.g002

In this model, the number of possible topics T is fixed. Three continuous random variables, $\theta$, $\phi$, and $\psi$, are involved in this model. For a given set of documents, $D^{train}$, our aim is to estimate the posterior distribution of those continuous random variables. Our inference scheme is based on the observation that

$$
\begin{aligned}
P(\theta,\phi,\psi|D^{train},\alpha,\beta,\mu) = \\
\sum_{z,x} P(\theta,\phi,\psi|z,x,D^{train},\alpha,\beta,\mu) P(z,x|D^{train},\alpha,\beta,\mu)
\end{aligned}
\tag{1}
$$

In the training process, an empirical sample-based estimation of $P(z,x|D^{train},\alpha,\beta,\mu)$ is first obtained using Gibbs sampling.

$$
\begin{aligned}
&P(z_{di},x_{di}|z_{-di},x_{-di},w,j,\alpha,\beta,\mu) \propto \\
&\frac{m_{x_{di}z_{di}}^{-di}+\alpha_{z_{di}}}{\sum_z (m_{x_{di}z}^{-di}+\alpha_z)} \frac{n_{z_{di}w_{di}}^{-di}+\beta_{w_{di}}}{\sum_{w_v}(n_{z_{di}w_v}^{-di}+\beta_{w_v})} \frac{n_{z_{di}j_d}^{-d}+\mu_{j_d}}{\sum_j (n_{z_{di}j}^{-d}+\mu_j)}
\end{aligned}
\tag{2}
$$

where the superscript $-di$ denotes a quantity, excluding the current instance (e.g., the $di$-th word token in the $d$-th paper). After Gibbs sampling, the probability of a word given a topic $\phi$, the probability of a journal given a topic $\psi$, and the probability of a topic given a bio-term $\theta$ can be estimated as follows:

$$
\phi_{zw_{di}} = \frac{n_{z_{di}w_{di}}^{-di}+\beta_{w_{di}}}{\sum_{w_v}(n_{z_{di}w_v}^{-di}+\beta_{w_v})}
\tag{3}
$$

$$
\psi_{z_j|d} = \frac{n_{z_{di}j_d}^{-d}+\mu_{j_d}}{\sum_j (n_{z_{di}j}^{-d}+\mu_j)}
\tag{4}
$$

$$
\theta_{xz} = \frac{m_{xz}+\alpha_z}{\sum_{z'} (m_{xz'}+\alpha_{z'})}
\tag{5}
$$

With the estimated continuous random variables, $\theta$, $\phi$, and $\psi$, we can identify the information content of bio-terms, and find association among bio-terms.

**Bio-term Entropy over Topics.** In information theory, entropy is a measure of the uncertainty associated with a random variable. It is also a measure of the average information content. In our Bio-LDA model, we can compute the bio-term entropies over topics as shown in equation 6, which indicates that bio-terms tend to address a single topic or cover multiple topics. The higher the entropy is, the more diverse the bio-term is over topics.

$$
Entropy(b_i) = -\sum_{z=1}^{T} \theta_{b_i z} \log \theta_{b_i z}
\tag{6}
$$

**Semantic Association of Bio-Terms.** Kullback-Leibler divergence (KL divergence) is a non-symmetric measure of the difference between two probability distributions. In our Bio-LDA model, we used the KL divergence as the non-symmetric distance measure for two bio-terms over topics, as shown in equation 7.

$$KL(b_i, b_j) = \sum_{z=1}^{T} \theta_{b_j z} \log \frac{\theta_{b_j z}}{\theta_{v_j z}} \qquad (7)$$

The symmetric distance measure of two bio-terms over topics is the sum of two non-symmetric distances, as shown in equation 8.

$$sKL(b_i, b_j) = \sum_{z=1}^{T} (\theta_{b_j z} \log \frac{\theta_{b_j z}}{\theta_{b_j z}} + \theta_{b_j z} \log \frac{\theta_{b_j z}}{\theta_{b_j z}}) \qquad (8)$$

In this study, the direction of the associations is not considered. We only focus on the association score calculated by the symmetric distances unless user specified to use the non-symmetric distances.

## Other implemented tools

In our Bio-LDA model, the association of two bio-terms in the literature can be measured use the KL-divergence. The smaller the score is, the stronger the association is. This association score can combined with the pre-knowledge of bio-terms (i.e. Chem2-Bio2Rdf) for novel knowledge discovery.

**Association Predication.** Two Bio-terms can be associated if there is a path between them or two bio-terms have similar chemical or biological activities. The graph definition of three types of semantic association is shown in Figure 3 [26]. However, the number of association pairs is usually very large for the big network. The association score from bio-LDA model can be used to rank and select the most interest pairs from the candidate pool.

In Chem2Bio2RDF, there are eight kinds of relations: *compound-compound, compound-gene, compound-disease, compound-side effect, compound-pathway, gene-gene, gene-disease,* and *gene-pathway*. For a given source, it is quite easy to find its associated target with a given type in Chem2Bio2RDF. For instance, finding compounds that target genes can be done by finding the direct relation among compound-gene pairs. However, users are usually not only interested in those already known links but also want to get information about the possible indirect links. Indirect relations through an intermediary also offer an opportunity to find linked compound-gene relations (ρ-path association). The relations can then be validated using the calculated association scores from the Bio-LDA topic model. For instance, in order to find the possible in-directed linkage for a given gene-compound pair, we can look up the four extended associations in Chem2Bio2RDF: *gene-disease-compound, gene-compound-compound, gene-pathway-compound,* and *gene-gene-compound,* and then compute the association scores for the outputs from Chem2Bio2RDF. A valid extended association is defined as following:

$$\left.\begin{array}{r} sKL(b_1, b_2) \leq c_T \\ Association(b_1, b_2) \end{array}\right\} \Rightarrow \hat{R}_T(b_1, b_2) \qquad (9)$$

where *association($b_1, b_2$)* indicates that possible semantic ρ-associations from Chem2Bio2RDF, *sKL($b_1, b_2$)* is the association score calculated using the Bio-LDA model described in section 3.2.2.

**Association Search.** In the area of network analysis, the task of association search can be formalized as a task of path find in graph. In our study, we are given a semantic network (e.g., Chem2Bio2RDF), which can be represented as a graph $G = (V, E)$, where $v \in V$ represents an entity (e.g., drug and gene) in the network; $e'_{ij} \in E$ represents a relationship with property $r$ (e.g., drug-target interaction) between entities $v_i$ and $v_j$; the relationship can be directional or bi-directional; the goal of association search is to find relationship sequences from $v_i$ to $v_j$. We assume that no entity will appear on a given association more than one time. We then define the process of association search from one entity to the other as: *Given an association query $(v_i, v_j)$, where $v_i$ denotes the source entity and $v_j$ denotes the target entity. Association search is to find possible associations $\{\alpha_k(v_i, v_j)\}$ from $v_i$ to $v_j$.*

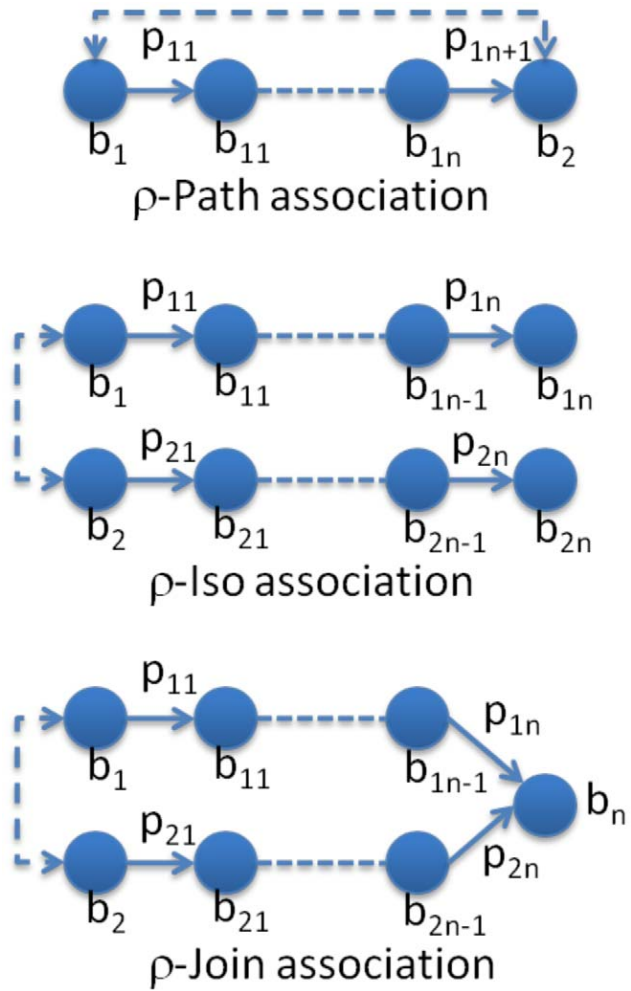

**Figure 3. Semantic Association.**
doi:10.1371/journal.pone.0017243.g003

There are two subtasks along with association search: finding possible associations between two bio-terms and ranking the associations. In this work, we formalize the association search problem as that of near-shortest associations search. We used a two-stage approach for finding the near-shortest associations for an association query $(v_i, v_j)$:

1. Shortest association finding. It aims at finding the shortest associations from all entities $v \in V \backslash v_j$ in the network to the target entity $v_j$ (including the shortest association from $v_i$ to $v_j$ with length $L_{min}$). In a graph, the shortest path between two nodes can be found using a heap-based Dijkstra algorithm to quickly find the shortest associations that can achieve a complexity of $O(n \log n)$.

2. Near-shortest associations finding. Based on the length of shortest association $L_{min}$ and a pre-defined parameter $\beta$, the algorithm requires enumeration of all associations that are less than $(1+\beta)L_{min}$ by a depth-first search. We constrain the length of an association to be less than a pre-defined threshold. This length restriction can reduce the computational cost.

The obtained associations are then ranked according to the accumulated KL-divergence scores obtained from the Bio-LDA model (*eq. 8*).

**Connectivity Map Generation.** The concept of molecular connectivity maps is gaining an increased popularity in systems chemical biology [27], which can help researchers to study and compare the molecular therapeutic/toxicology profiles of many candidate drugs. In this work, we proposed a computational approach to build interest-specific connectivity maps, i.e. build disease-specific gene-drug connectivity maps, based on both the genomic data sources and the literature resources. The input query for the connectivity map is $\langle(v_i, t), t_1, t_2\rangle$, where $(v_i, t)$ is the specified bio-term and its type (e.g. Alzheimer's Disease, Disease), $t_1$ and $t_2$ are the bio-term types that formed the connectivity maps (e.g., drug and gene). In this study, the candidate bio-terms for $t_1$ are identified and refined based on the genomic data sources in Chem2Bio2Rdf and the candidate bio-terms for $t_2$ are those that could interact to the candidate bio-terms for $t_1$. Their connection scores are given: the linkage in Chem2Bio2Rdf, the association score calculated based on the Bio-LDA model (equation. 8). Here, we used the disease-specific gene-drug connectivity map as an example to show the process:

1. Specify a disease
2. Identify the genes that related to the given disease from prior knowledge (i.e. Chem2Bio2Rdf).
3. Expand the genes based on the gene-gene interaction. This step can be ignored if user does not want to count the protein-protein interaction.
4. The genes identified from step 2 and 3 are combined and re-ranked.
5. Find the drugs that can target one or more genes from the gene set given by step 4. The drugs are ranked based on an accumulated score of the importance of the targeted genes.
6. Calculate the association score based on the Bio-LDA model for the gene set from step 4 and the drug set from set 5 to form the connectivity maps.

## Results

### Analyzing the Bio-LDA Model Results

In our experiments, we applied the Bio-LDA model to 336,899 MEDLINE abstracts ($\sim$ 330M in size) published in 2009, which contains 308686 words, 13338 extracted bio-terms (only drug, gene, disease are considered in this experiment), and 4450

**Table 2.** Representations for selected topics.

| Topic 13 | | | Topic 14 | | |
|---|---|---|---|---|---|
| **Word** | | **Prob** | **Word** | | **Prob** |
| patient | | 0.0177 | patient | | 0.0231 |
| transplant | | 0.0149 | liver | | 0.0129 |
| platelet | | 0.0074 | hepat | | 0.0126 |
| studi | | 0.0066 | diseas | | 0.008 |
| group | | 0.0063 | studi | | 0.007 |
| donor | | 0.0058 | treatment | | 0.0063 |
| factor | | 0.0056 | result | | 0.0059 |
| risk | | 0.0054 | group | | 0.0057 |
| result | | 0.0053 | hcv | | 0.0056 |
| graft | | 0.0053 | associ | | 0.0052 |
| **Bio-Terms** | **type** | **Prob** | **Bio-Terms** | **type** | **Prob** |
| Thrombosis | DISEASE | 0.0855 | Hepatitis C | DISEASE | 0.0883 |
| Venous Thromboembolism | DISEASE | 0.0449 | Colitis | DISEASE | 0.0784 |
| Heparin | DRUG | 0.0417 | Hepatitis B | DISEASE | 0.0511 |
| Tacrolimus | DRUG | 0.0402 | Hepatitis | DISEASE | 0.0467 |
| Cyclosporine | DRUG | 0.0338 | Fibrosis | DISEASE | 0.0383 |
| VWF | GENE | 0.0335 | Fatty Liver | DISEASE | 0.0274 |
| Thrombocytopenia | DISEASE | 0.0274 | Ribavirin | DRUG | 0.0258 |
| Mycophenolate mofetil | DRUG | 0.0259 | Liver Cirrhosis | DISEASE | 0.0236 |
| IMPACT | GENE | 0.0225 | Gastroesophageal Reflux | DISEASE | 0.0229 |
| ABO | GENE | 0.0223 | Irritable Bowel Syndrome | DISEASE | 0.0222 |
| **Journal** | | **Prob** | **Journal** | | **Prob** |
| Transplant. Proc. | | 0.0734 | Hepatology | | 0.064 |
| Transplantation | | 0.0721 | World J. Gastroenterol. | | 0.0553 |
| Thromb. Haemost. | | 0.0431 | Am. J. Gastroenterol. | | 0.0532 |
| Thromb. Res. | | 0.0428 | Gastroenterology | | 0.0477 |
| Transfusion | | 0.0412 | Liver Int. | | 0.0394 |

doi:10.1371/journal.pone.0017243.t002

journals, using a default maximum association path length of 3. The output includes the estimated parameters, $\theta$, $\phi$, and $\psi$, and the top term lists of words, bio-terms and journals for each topic. We created different models with 50, 100 and 200 topics. No significant improvement was found with increased numbers of topics. Thus, we used 50-topics model to optimize efficiency.

Examples of two topics (out of 50 topics in total) with the top 10 representative words, top 10 associated bio-terms, and top 5 related journals are listed in Table 2. The Bio-LDA model provides an unsupervised method for extracting an interpretable representation from a collection of documents. As shown in the Table 2, the topic 13 is related to organ transplant and all of the highest probability BioTerms for topic 14 are highly related to liver disease (hepatitis). Our Bio-LDA model used the bio-terms, journal information and the word information to characterize the topic providing a better representation of topics than the simple LDA model, which only can provide the word representation

Table 3 shows a table with the most associated topics for 3 of 13,338 possible bio-terms. The first bio-term, tuberculosis, is an infectious lung disease caused by various strains of mycobacterium. Topic 21 is the majority topic associated with tuberculosis with a conditional probability of 0.8024 and very low probabilities for all other topics. TNF, tumor necrosis factor, are almost equally distributed to topic 33 and 38, with probability of 0.5203 and 0.4149. The last bio-term, cholesterol, is a waxy steroid metabolite found in the cell membranes and transported in the blood plasma of all animals. The topics associated with this drug are quite intuitive.

The word representations of topics provide an overview of the published literature. Research trends over time could be discovered by applying the Bio-LDA topic model on different years individually and comparing results (looking for emerging topics).

## Comparing the Bio-LDA and LDA models

We chose the top 20 representative words for all 50 topics and computed the word frequency based on the Bio-LDA model and the general LDA model. As shown in Table 4a, 635 distinct words are used to represent topics for LDA model and 354 distinct words for Bio-LDA model. 462 words only appeared once in the top 20 topic words for the LDA model and 234 words for the Bio-LDA model. It is shown that Bio-LDA tended to use fewer words to represent topics.

In order to compare the output, we mapped the topics generated using the LDA model with the topics generated using the Bio-LDA model. To map the Bio-LDA model to the LDA model, we searched the top 20 words for all topics in the LDA model for each topic in the Bio-LDA model. The topic with the highest number of shared words is considered as the mapped topic in the LDA model. As shown in Table 4b, the 50 topics in the Bio-LDA are mapped to 25 topics in the LDA. Only 17 topics in the Bio-LDA model can mapped to unique topics in the LDA model. The reverse mapping gave better performance. The 50 topics in the LDA model can be mapped to 39 topics in the Bio-LDA model. About 30 topics have unique mappings. Table 5 shows three mapping examples of mapping LDA to Bio-LDA. Topic 30 in the LDA model is mapped to topic 25 in Bio-LDA model; topic 41 is mapped to topic 33, and topic 25 is mapped to topic 38. There are 11 common words for each mapping.

## Identification of Bio-Term Relationships within Topics

In the biomedical literature, bio-terms (drug names, gene names, diseases, etc.) play an important role in determining the topics. The Bio-LDA model makes direct use of bio-terms to improve the overall topic generation and word association. As shown in Table 6, only 55 words (5 drugs, 17 genes, and 33 diseases) are bio-terms among the 635 unique words (Table 7a) generated from the top 20 words of the 50 topics in the LDA model contains only. There are 66 bio-terms (9 drugs, 17 genes, and 40 diseases) among the 354 unique words in the Bio-LDA model. The top 20 bio-terms associated with topics are also output for our Bio-Terms. Thus, significant number of bio-terms can be identified in the Bio-LDA model. As shown below, 663 distinct bio-terms, including 145 drugs, 150 genes, and 368 diseases are identified.

We assume that there exists a weak topic related association if the bio-terms are in the top list of a topic. Figure 4 illustrates an association network drawn from the top 5 terms from 6 selected

**Table 3.** Top topics for the selected bio-terms.

| **BioTerm = Tuberculosis (Disease)** | | |
|---|---|---|
| P(z|b) | Topic | Words |
| 0.8024 | 21 | infect, hiv, patient, vaccin, studi, case, tuberculosi, result, year, risk |
| 0.0841 | 12 | gene, protein, cell, express, strain, infect, pathogen, these, host, respons |
| 0.0594 | 29 | protein, bind, activ, structur, cell, domain, interact, these, membran, site |
| **BioTerm = TNF (Gene)** | | |
| P(z|b) | Topic | Words |
| 0.5203 | 33 | cell, express, inflamm, activ, inflammatori, induc, increas, alpha, effect, level |
| 0.4149 | 38 | cell, express, activ, immun, respons, induc, mice, cd4, receptor, these |
| 0.0341 | 30 | cell, activ, effect, induc, rat, studi, increas, oxid, level, express |
| **BioTerm = Cholesterol (Drug)** | | |
| P(z|b) | Topic | Words |
| 0.3314 | 31 | weight, obes, studi, associ, risk, women, children, group, bodi, increas |
| 0.2926 | 33 | cell, express, inflamm, activ, inflammatori, induc, increas, alpha, effect, level |
| 0.1072 | 36 | insulin, diabet, patient, glucos, level, studi, type, increas, associ, result |

**Table 4.** a) Frequency word sets of LDA model and Bio-LDA model. b) Mappings between Bio-LDA model and LDA model.

| Bin | LDA | Bio-LDA | Bin | BioLDA2LDA | LDA2BioLDA |
|-----|-----|---------|-----|------------|------------|
| 1 | 462 | 234 | 1 | 17 | 30 |
| 2 | 100 | 52 | 2 | 2 | 7 |
| 3 | 32 | 22 | 3 | 2 | 2 |
| 4 | 19 | 11 | 4 | 2 | 0 |
| 5 | 8 | 2 | 5 | 0 | 0 |
| 6 | 4 | 0 | 6 | 0 | 0 |
| 7 | 2 | 4 | 7 | 1 | 0 |
| 8 | 1 | 4 | 8 | 1 | 0 |
| 9 | 0 | 4 | | | |
| 10 | 4 | 2 | | | |
| >10 | 3 | 19 | | | |
| SUM | 635 | 354 | SUM | 25 | 39 |
| | (a) | | | (b) | |

doi:10.1371/journal.pone.0017243.t004

**Table 6.** Bio-terms associated with topics.

| Top 20 | LDA | Bio-LDA | |
|--------|-----|---------|---|
| | words | words | bio-terms |
| Drug | 5 | 9 | 145 |
| Gene | 17 | 17 | 150 |
| Disease | 33 | 40 | 368 |
| bio-terms | 55 | 66 | 663 |

doi:10.1371/journal.pone.0017243.t006

which are not direct enough to be in a dataset. For instance, our Bio-LDA model suggested that there might be protein protein-protein interactions between CCND1 and EGFR, since both of them are important targets in the tumor related diseases: Carcinoma and Melanoma. It also suggested that EGFR may also be target for Melanoma, although the linked database does not mention it.

### Discovery of Bio-Term Associations

In traditional text mining, bio-term association is usually calculated based on the literature co-occurrence of those two terms (Li J, et. al, 2008):

$$\Theta(b_i, b_j) = \ln(df(b_i, b_j) * N + \lambda) - \ln(df(b_i) * df(b_j) + \lambda) \quad (10)$$

Here, $df(b_i)$ and $df(b_j)$ are the number of documents in which bio-terms $b_i$ and $b_j$ are mentioned, respectively, $df(b_i, b_j)$ are the total number of documents in which both bio-terms are co-mentioned in the same document. $N$ is the size of the document collection. $\lambda$ is a small constant ($\lambda = 1$ here) introduced to avoid out-of-bound errors if any of $df(b_i, b_j)$, $df(b_i)$ or $df(b_j)$ values are 0. The $\Theta(b_i, b_j)$ representing the connections between the two bio-terms. It is positive when the potential pairs are over-represented and negative when the pairs are under-represented. The higher the $\Theta(b_i, b_j)$ is, the more significant the two bio-terms are connected.

However, a big limitation of this method is that it cannot detect association between two bio-terms if they are not involved in the same document. For example, the HTR1A and HTR2A both do not appear in same abstracts as Venlafaxine based on the PubMed collection. So the calculated association scores are negative as shown in table 7, which means there shouldn't be any association between Venlafaxine and HTR1A or HTR2A. However, we known that Venlafaxine is used in the treatment of mental disorder, e.g. depressive disorder and anxiety disorder. HTR1A and HTR2A have also been studied in relation to mental disorders. So, in reality, there must be certain association between Venlafaxine and HTR1A and HTR2A.

To cover the drawbacks of this co-occurrence based method, a better association approach based on the Bio-LDA topic model is used. In the Bio-LDA model, venlafaxine, HTR1A and HTR2A are all signed to topic 10, which focus on research on mental diseases (top 5 word representation of topic 10 are patient, studi, depress, schizophrenia, and treatment). The calculated association score between venlafaxine and HTR1A is quite small, indicating a very strong association between venlafaxine and HTR1A. It is also in agreement with our previous explanation.

topics. The colors of lines are used to present various topics. Solid lines indicate the generated relationships are proofed by Chem2Bio2RDF (i.e. there is a relationship in one of the Chem2Bio2RDF datasets that confirms the literature relationship). The dashed lines indicates the generated relationships have not been found in Chem2Bio2Rdf, perhaps indicating very recent findings which are not yet encoded in databases, or associations

**Table 5.** Compare word representation of topics in the Bio-LDA model to topics in the LDA model.

| 30<->25 | | 41<->33 | | 25<->38 | |
|---------|---------|---------|---------|---------|---------|
| LDA | Bio-LDA | LDA | Bio-LDA | LDA | Bio-LDA |
| cell | cell | alpha | cell | cell | cell |
| induc | cancer | factor | express | respons | express |
| apoptosi | express | inflammatori | inflamm | immun | activ |
| line | tumor | induc | activ | antibodi | immun |
| effect | activ | beta | inflammatori | specif | respons |
| human | gene | endotheli | induc | antigen | induc |
| inhibit | protein | increas | increas | anti | mice |
| death | induc | inflamm | alpha | gamma | cd4 |
| growth | human | express | effect | lymphocyt | receptor |
| prolifer | growth | activ | level | ifn | these |
| activ | inhibit | effect | mice | cd4 | cytokin |
| vitro | studi | tnf | protein | induc | human |
| p53 | effect | vascular | factor | product | specif |
| result | result | growth | studi | activ | regul |
| increas | associ | role | tnf | cytokin | function |
| cycl | line | cytokin | cholesterol | human | antigen |
| caspas | these | macrophag | result | against | infect |
| treatment | apoptosi | mmp | role | receptor | role |
| tumor | breast | tissu | receptor | cd8 | mediat |
| vivo | regul | matrix | these | system | signal |

doi:10.1371/journal.pone.0017243.t005

**Table 7.** Calculated association score for Venlafaxine and HTR1A, HTR2A.

| Bio-terms | Co-occurrence | Bio-LDA |
|---|---|---|
| Venlafaxine ~ HTR1A | −11.76 | 0.34 |
| Venlafaxine ~ HTR2A | −12.72 | 4.0 |

doi:10.1371/journal.pone.0017243.t007

In order to get a quantitative measurement of the goodness of our Bio-LDA model in discovering the bio-term associations, the bio-term pairs in Chem2Bio2Rdf are used as the gold standard. As shown in table 8, only few bio-term pairs are identified using the co-occurrence method. The KL-divergence method based on the Bio-LDA model can identify a much larger number of association pairs. The cut-off for co-occurrence method is 0 and the cut-off for Bio-LDA model is 5.

### Identifying Potential Drugs for a Target

As discussed in section 3.3.1, we can generate bio-term associations by combining the linked data resources (i.e. Chem2Bio2Rdf) and literature resources (Bio-LDA model). To illustrate, we investigated drugs that target the abelson murine leukemia viral oncogene homolog 1 (ABL1), which has been implicated in processes of cell differentiation, cell division, cell adhesion, and stress response. ABL1 is also known as a factor in chronic myeloid leukemia. As shown in Figure 5, five drugs, cisplatin, adenesine triphosphate, imatinib, dasatinib, and gefitinib, target ABL1 according to Drugbank (accounting for the solid lines in the diagram). We make predictions about drugs that may target ABL1 from our model generated via the gene-disease-drug association which must satisfy two conditions:

1) There exists a gene-disease-drug path in Chem2Bio2RDF
2) The calculated association score for the gene-drug pair should be less than a certain threshold.

The association scores are computed using the Bio-LDA model with 50 topics on the most recent 336899 abstracts published recently. The association score based on the Bio-LDA model are given by equation 8, which is also known as the symmetric KL divergence. We used the score not larger than 5 as the threshold. Usually, there exist multiple gene-disease-drug paths in Chem2-Bio2Rdf for a given gene-drug pair. The accumulated score of each pair in the path is used to rank the possible paths and only the one with the most significant score will be shown in the network. The diseases, leukemia, myeloma and neoplasm, are the most significant diseases that associate the gene with drugs. Figure 5 shows the generated network using the Bio-LDA model. 15 drugs are suggested by the Bio-LDA model. Similar to the directly linked five drugs, those predicted drugs are all chemotherapy related drugs. The diseases, leukemia and multiple myeloma, are also highly associated with the ABL1 based on our analysis.

### Investigating Drug Polypharmacology

In drug discovery, a major question is how to find drug candidates for a targeted disease. Since approximately 35% of known drugs have more than one target, the efficacy of many drugs is increasingly thought to come from their effect on multiple targets, which is known as polypharmacology. Based on this assumption, drug candidates can be identified from compounds, which have the same multiple targets as a marketed drug. Thus, the question of how to find drug candidate for a therapy can be formulated as a query in our system: find all drug-like compounds that share at least two targets with the drug that used for the therapy. For example, if a user wants to find some drug candidates for inflammatory and autoimmune conditions, such as rheumatoid arthritis, he can start with the typical drug, dexamethasone, and then search for the compounds that active the similar targets with a activity score greater than 50 (activity score 0–100). The graphical representation of an example of the query process is shown in the following Figure 6.

To further understand the relation between the given drug, dexamethasone, and the found compound, hydrocortisone, we

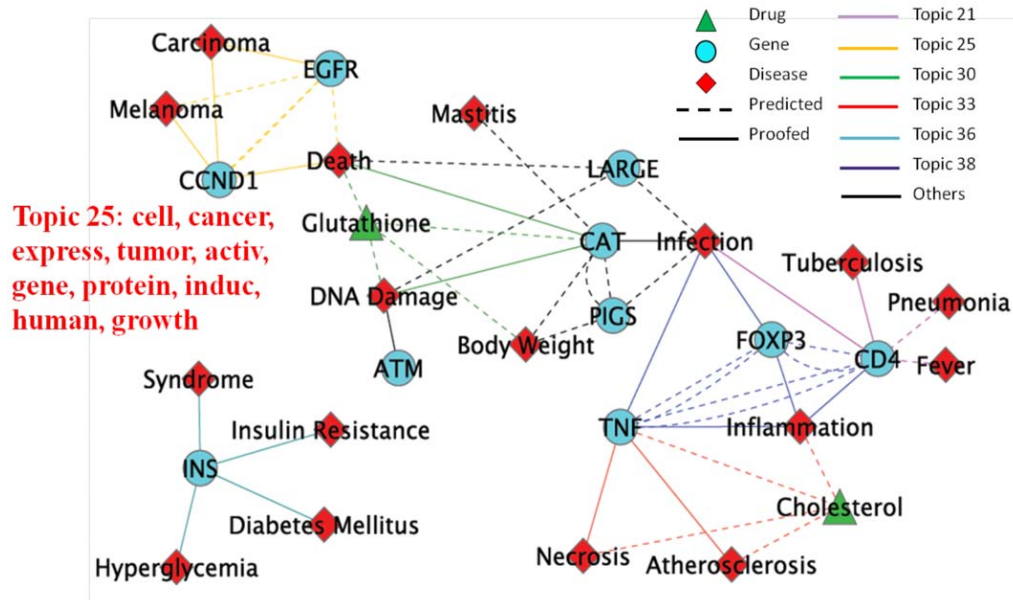

**Figure 4. An illustration of bio-relationships generated from selected topics.**
doi:10.1371/journal.pone.0017243.g004

**Table 8.** Comparing the co-occurrence method and the Bio-LDA in identifying associated bio-terms.

| Bio-terms | Chem2Bio2Rdf | Co-occurrence | Bio-LDA |
|---|---|---|---|
| Disease ~ Gene | 412117 | 266 | 14895 |
| Disease ~ Drug | 1490 | 20 | 228 |
| Gene ~ Drug | 5047 | 28 | 355 |
| Gene ~ Gene | 7593 | 13 | 1282 |

doi:10.1371/journal.pone.0017243.t008

apply the semantic association search proposed in section 2.3.4 within our Chem2Bio2Rdf. 47 near-shortest paths are found from hydrocortisone to dexamethasone including 5 types, drug-gene-drug, drug-gene-drug, drug-disease-drug, drug-gene-gene-drug, drug-gene-disease-drug, and drug-disease-gene-drug. Those near-shortest paths are then ranked based on the association scores calculated using equation 10. The top 10 paths and the association scores for each pairs (based on 50 topics) are shown in Figure 7. As shown in graph, three similar gene targets, NR3C1, ANXA1 and NOS2 are shared by both dexamethasone and hydrocortisone. Among those paths, five paths are associated with NR3C1, Glucocorticoid receptor, which indicates its significant role in understanding pharmacokinetic of drugs.

Table 9 shows the entropy of the two drugs and three gene targets calculated based on the Bio-LDA model with 50, 100 and 200 topics using the recent 336,899 MEDLINE abstracts, which contain 13,338 identical bio-terms. Here $n$ represents the number of abstracts that contain the given bio-terms in the literature set. Dexamethasone is a more effective drug when compared to Hydrocortisone, since it is involved in 742 more abstracts and has higher entropies. This makes sense from a biological point of view as dexamethasone is 40 times more potent than hydrocortisone. Table 10 shows the symmetric KL divergence for pairs of bio-terms in this use case, and $n$ shows the number of co-occurrence of the given bio-term pair. Hydrocortisone and dexamethasone co-occurred in 17 abstracts and have lower KL divergence.

Hydrocortisone and dexamethasone target genes NR3C1, ANXA1, and NOS2. Thus what do the entropy and KL divergence indicate about the features of those two paths? For different number of topics (T = 200, T = 100, and T = 50), Table 9 shows that these ascending order of average entropy for the three genes is: ANXA1<NR3C1<NOS2, suggesting NOS2 tends to be involved with more topics while ANXA1 tends to be associated with less topics. Thus the path between the two drugs with ANXA1 is more focused and specific, which intuitively conveys more meaning. This makes sense as hydrocortisone and dexamethasone are involved in de novo synthesis of ANXA1 gene (Mulla, A, et. al, 2005). Thus the three paths involved with the three genes can be ranked according to their semantic specificity as: path with ANXA1>path with NOS2>path with NR3C1. Moreover, the smaller the KL divergence of the values of the path is, the more semantically relevant are the nodes and edges along the path. Table 10 shows that the entities and relationships along the path through NR3C1 are the most relevant to each other of the three paths. Combining entropy and KL divergence, the path with ANXA1 is more favorable in specific research and the path with NR3C1 is more favorable in general research.

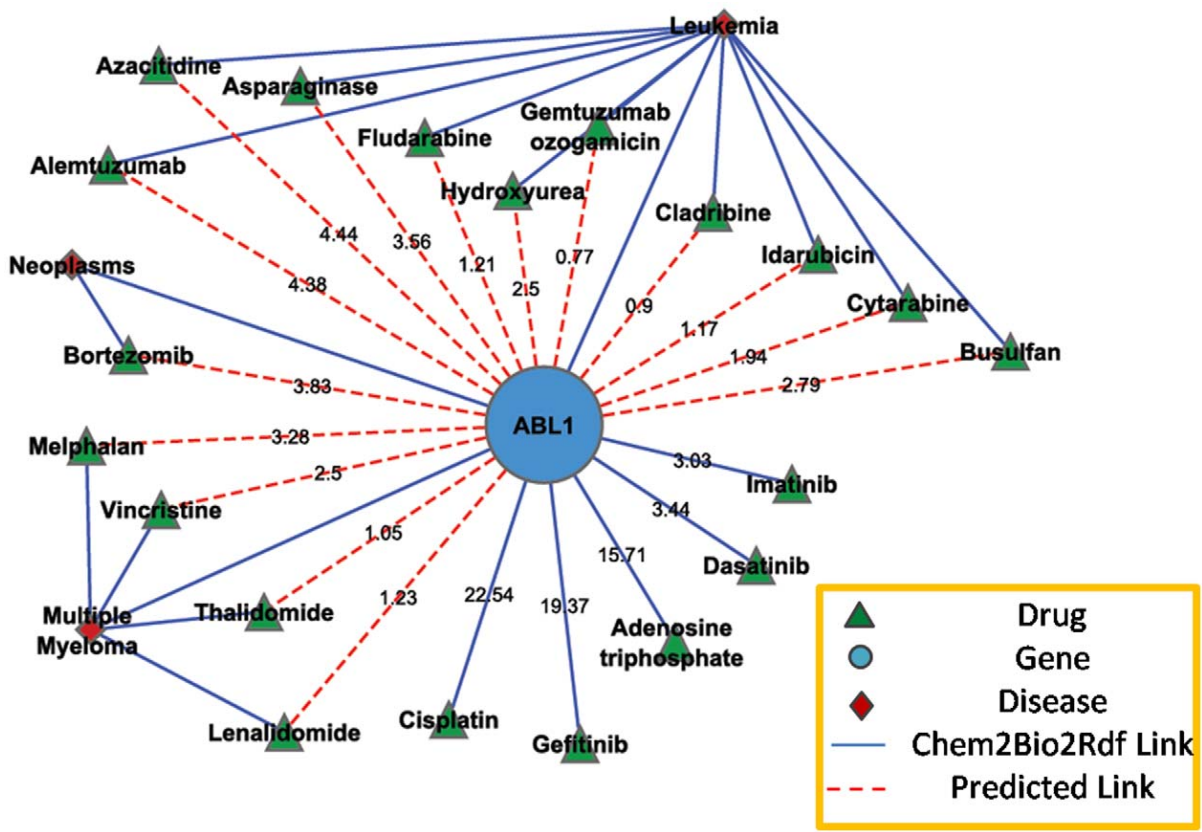

**Figure 5. The association network for Tyrosine-protein kinase ABL1 based on the Bio-LDA model.**
doi:10.1371/journal.pone.0017243.g005

  

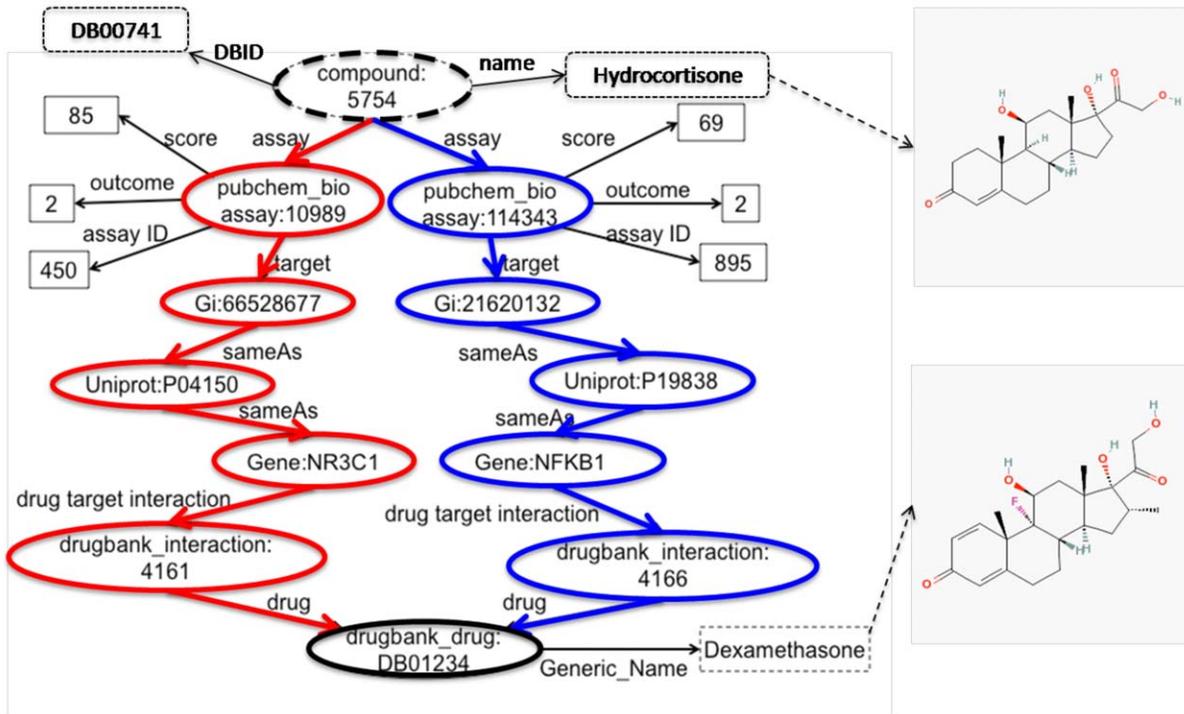

**Figure 6. Graphic representation of the SPARQL query for finding the compound similar to Dexamethasone.**
doi:10.1371/journal.pone.0017243.g006

## Building a Disease-Specific Drug-Protein Connectivity Map

The molecular connectivity map shows how the expression level of genes change in response to different drug compound perturbations, which enables researchers to compare the molecular therapeutic/toxicological profiles of many candidate drugs or drug target genes, therefore improving the chance of developing high quality drugs and reducing drug development time. In this

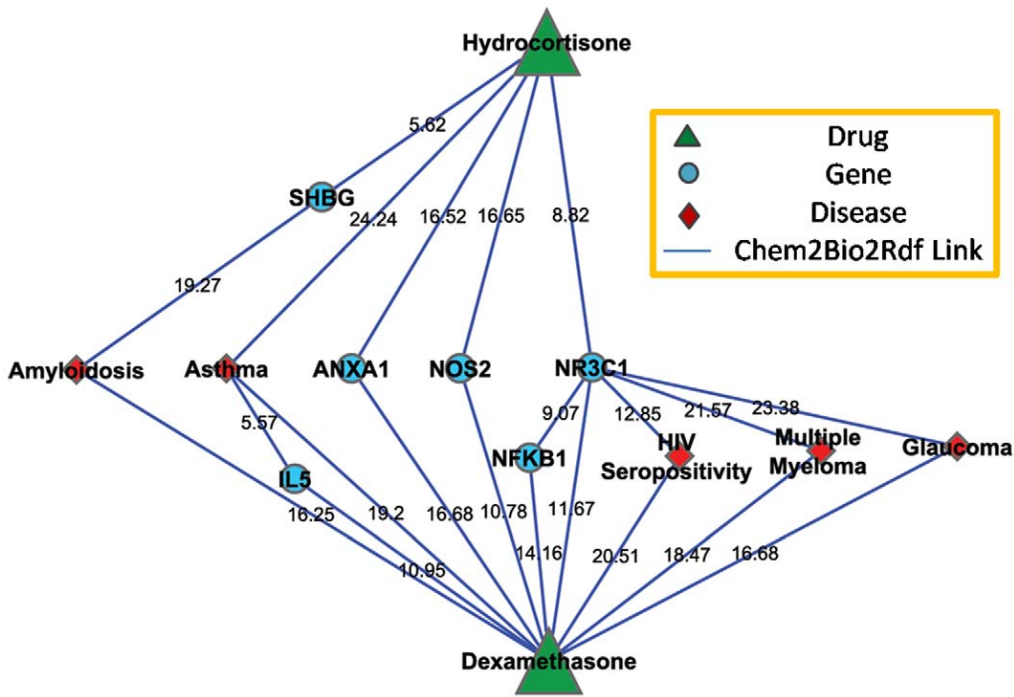

**Figure 7. The top 10 paths obtained between Hydrocortisone and Dexamethasone.**
doi:10.1371/journal.pone.0017243.g007

**Table 9.** Bio-term entropies for nodes shown in the top 3 paths.

| Bio-terms name | Bio-terms Identify | Type | n | T = 200 | T = 100 | T = 50 | Average |
|---|---|---|---|---|---|---|---|
| Hydrocortisone | DB00741 | Drug | 139 | 2.558 | 1.880 | 2.454 | 2.297 |
| Dexamethasone | DB01234 | Drug | 881 | 4.292 | 3.754 | 3.484 | 3.843 |
| ANXA1 | P04083 | Gene | 23 | 2.266 | 1.631 | 1.365 | 1.754 |
| NR3C1 | P04150 | Gene | 16 | 2.123 | 2.840 | 2.486 | 2.483 |
| NOS2 | P35228 | Gene | 40 | 2.824 | 2.833 | 2.598 | 2.752 |

doi:10.1371/journal.pone.0017243.t009

**Table 10.** Symmetric KL divergence for the top 3 paths.

| Bio-term semantic associations | T = 200 | T = 100 | T = 50 | Average |
|---|---|---|---|---|
| Hydrocortisone ~ NR3C1 ~ Dexamethasone | 29.96 | 21.55 | 20.49 | 24.00 |
| Hydrocortisone ~ NOS2 ~ Dexamethasone | 35.40 | 31.00 | 27.42 | 31.28 |
| Hydrocortisone ~ ANXA1~ Dexamethasone | 43.39 | 40.31 | 33.20 | 38.97 |

doi:10.1371/journal.pone.0017243.t010

study, we use a novel method to compute the high-coverage disease-specific drug-gene connectivity maps, by integrating chemogenomic sources (i.e. Chem2Bio2Rdf) with literature from our Bio-LDA model. The purpose of the connectivity maps is finding novel therapeutic uses of old drugs, also known as drug repositioning.

Using Alzheimer's disease (AD) as an example, the gene list is created by searching for AD-related genes from our linked data (Chem2Bio2Rdf). 88 genes were identifies. 382 drugs are selected based on the drug-gene interaction. The gene list can be expanded to 13998, and the drug list can expand to 1898 if the gene-gene interaction is involved. Our research shows that the top ranks of the expanded genes and drugs are close to the lists without extension. Thus, we do not consider gene-gene interaction unless user specified to speed up the calculation.

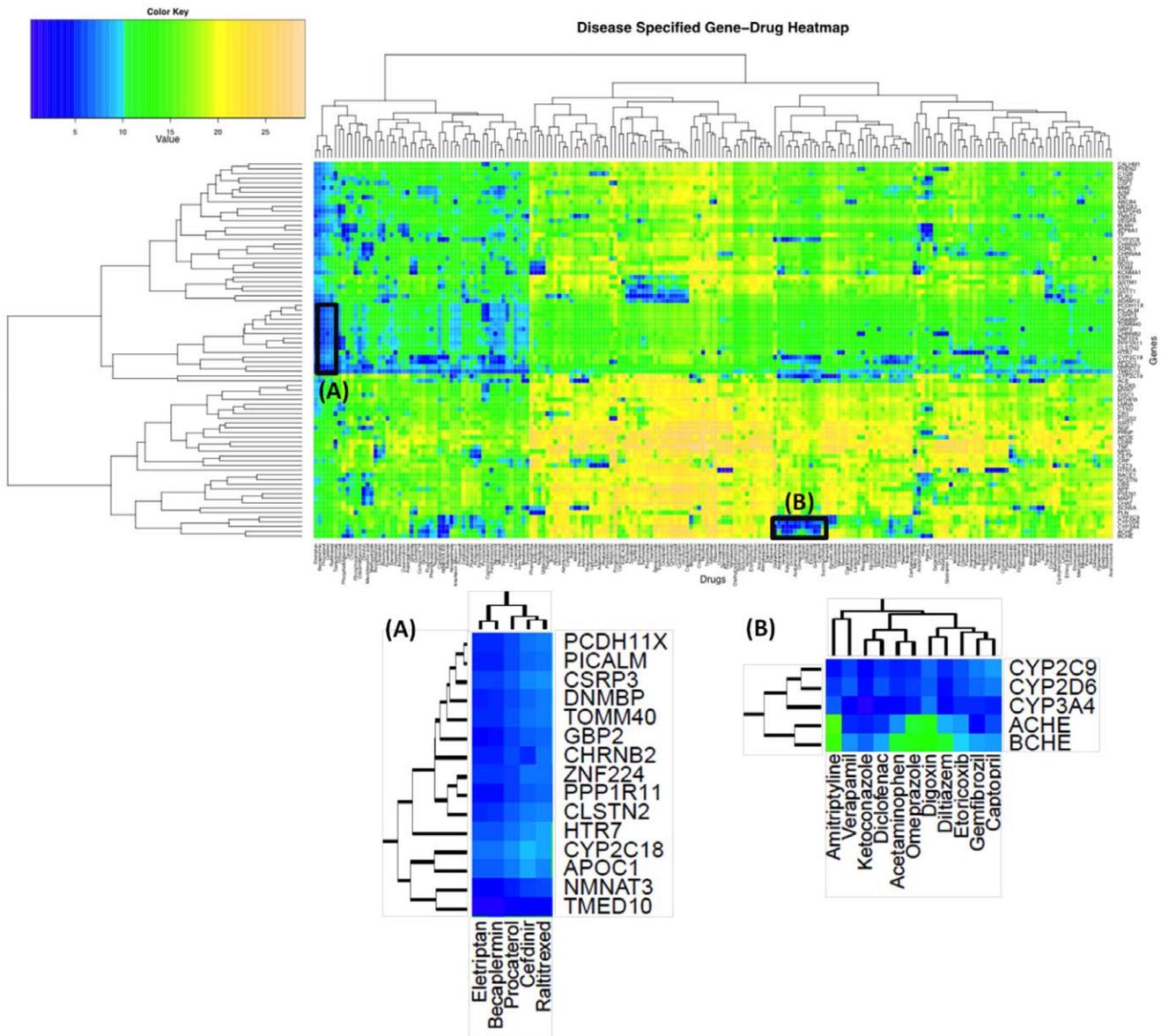

**Figure 8. A connectivity map linking AD-related genes to significant drugs, highlighting two areas (A) and (B).**
doi:10.1371/journal.pone.0017243.g008

The connectivity scores are calculated using the Bio-LDA association scores. Figure 8 shows the AD-related drug-protein connectivity map. The x-dimension represents drugs and the y-dimension represents genes. Hierarchical clustering of drugs and genes is performed use their Euclidean distances. The color intensity for each cell is drawn in proportion to the connectivity score as shown in the heatmap legend. In the Bio-LDA model, the connectivity scores indicate the distance between the gene-drug pair. The smaller the score is, the more significant the relationship is. The cells with purple color indicate the significant interactions related to Alzheimer's disease. From the figure and zoom in boxes, we can study the genes and drugs highly related to Alzheimer's disease. For example, the CYP family is known to be highly associated with AD. The discovered drug, Ketoconazole, may affect some AD drug metabolism, such as Donepezil. The Diclofenac is a non-steroidal anti-inflammatory drug (NSAID). Research shows the NSAIDs may prevent the development of AD if given daily in small doses during many years.

## Discussion

In this paper, we describe the architecture and main features of the Bio-LDA model. Three applications, association prediction, association search, and connectivity map generation, are presented which we believe are useful for biomedical and drug discovery applications, especially when combining the Bio-LDA model with a pre-knowledge network, i.e. Chem2Bio2Rdf. We believe these experiments demonstrate great value in performing this kind of analysis for enhancing biological knowledge.

We demonstrate how Bio-LDA, in contrast to natural language processing methods, can automatically derive a collection of topics of related biological terms that map to clearly understandable biological themes, and which allow the complexity of topics addressed in individual papers to be represented by probabilities of association with topics. Further, individual bioterms can be associated with topics with a given level of probability, and through the KL Divergence measure, a distance between any two terms can be generated via their probabilities of association with topics. This opens up the possibility of using the method for ranking paths through the data, or for an alternate way of measuring degree of association between, for example, drugs and genes, or pathways and diseases.

Our examples indicate that the topics created using Bio-LDA are surprisingly succinct in identifying the bioterms associated with particular topic areas. Our comparison of Bio-LDA with a standard LDA model showed that the models created by BioLDA are distinctly different from standard LDA and indicate that the use of bioterms only is useful in defining crisp clusters. Further, our case studies pertaining to drug targets and drug polypharmacology and indicate that when combined with methods for finding paths between entities, highly relevant results can be obtained (for example, finding potential drugs for a target or compounds with similar polypharmacology). Finally we show how Bio-LDA can be used to increase the utility of molecular connectivity approaches such as heatmaps.

Our experiment used 336899 recent MEDLINE abstracts. The performance of various LDA and extended LDA implementations is computationally expensive, motivating efforts to improve scalability. As the underlying algorithms for various implementations differ, the efforts to improve scalability have also differed [19,28,29]. In order to run our model in all 18 authors in PubMed, a scalable model, such as the parallel Bio-LDA model, is required. We plan to investigate the PLDA implementation of Wang et al with Bio-LDA, using the Map Reduce implementation [19]. By making larger collections available for analysis, we hope to expose better and more complex relations.

## Acknowledgments

We would like to thank Michel Dumontier and Glen Newton at Carleton University for their suggestions on this work.

## Author Contributions

Conceived and designed the experiments: HW YD JT XD BE JQ DW. Performed the experiments: HW YD JT XD BE JQ DW. Analyzed the data: HW YD JT XD BE JQ DW. Wrote the paper: HW YD DW.

## References

1. Muin M, Fontelo P, Ackerman M (2006) PubMed Interact: an interactive search application for MEDLINE/PubMed. AMIA Annual Symposium proceedings/ AMIA Symposium AMIA Symposium 1089.
2. Cohen KB, Hunter L (2004) Natural language processing and systems biology. Artificial intelligence and systems biology. pp 147–174.
3. Feldman R, Regev Y, Hurvitz E, Finkelstein-Landau M (2003) Mining the biomedical literature using semantic analysis and natural language processing techniques. 1: 69–80.
4. Blei D, Ng A, Jordan M (2003) Latent Dirichlet Allocation. Journal of Machine Learning Research 3: 993–1022.
5. Rosen-Zvi M, Griffiths T, Steyvers M, Smyth P (2004) The author-topic model for authors and documents; Banff, Canada: AUAI Press. pp 487–494.
6. Tang J, Zhang J, Yao L, Li J, Zhang L, et al. (2008) ArnetMiner:extraction and mining of academic social networks; Las Vegas, Nevada, USA: ACM. pp 990–998.
7. Wild DJ (2009) Mining large heterogeneous data sets in drug discovery. Expert Opinion on Drug Discovery 4: 995–1004.
8. Belleau Fo, Nolin M-A, Tourigny N, Rigault P, Morissette J (2008) Bio2RDF: towards a mashup to build bioinformatics knowledge systems. Journal of biomedical informatics 41: 706–716.
9. Chen B, Dong X, Jiao D, Wang H, Zhu Q, et al. (2010) Chem2Bio2RDF: a semantic framework for linking and data mining chemogenomic and systems chemical biology data. BMC Bioinformatics 11: 255.
10. Jentzsch AZJ, Hassanzadeh O, Cheung K, Samwald K, Andersson B (2009) Linking open drug data; Graz, Austria.

11. Hofmann T (2009) Probabilistic latent semantic indexing; Berkeley, California, United States: ACM. pp 50–57.
12. Wang X, McCallum A (2006) Topics over time: a non-Markov continuous-time model of topical trends; Philadelphia, PA, USA: ACM. pp 424–433.
13. Si X, Sun M (2009) Tag-LDA for Scalable Real-time Tag Recommendation. Journal of Computational Information Systems.
14. Xu H, Wang J, Hua X, Li S (2009) Tag refinement by regularized LDA; Beijing, China: ACM. pp 573–576.
15. Wang X, Grimson WEL, Westin C-F (2007) Tractography segmentation using a hierarchical Dirichlet processes mixture model.
16. Blei D, McAuliffe J (2010) Supervised Topic Models. Available: http://arxiv.org/abs/1003.0783v1, Accessed 2010 Jul 3.
17. Newman D, Asuncion A, Smyth P, Welling M (2007) Distributed inference for latent Dirichlet allocation. Neural Information Processing Systems (NIPS) 20: 1081–1088.
18. Steyvers M, Smyth P, Zvi M, Griffiths T (2004) Probabilistic author-topic models for information discovery; Seattle, WA, USA: ACM. pp 306–315.
19. Wang Y, Bai H, Stanton M, Chen W-Y, Chang EY (2009) PLDA: Parallel Latent Dirichlet Allocation for Large-Scale Applications; San Francisco, CA, USA: Springer-Verlag. pp 301–314.
20. Blei DM, Franks K, Jordan MI, Mian IS (2006) Statistical modeling of biomedical corpora: mining the Caenorhabditis Genetic Center Bibliography for genes related to life span. BMC Bioinformatics 7.
21. Zheng B, McLean D, Lu X (2006) Identifying biological concepts from a protein-related corpus with a probabilistic topic model. BMC Bioinformatics 7: 58–58.

22. Mörchen F, Dejori Mu, Fradkin D, Etienne J, Wachmann B, et al. (2008) Anticipating annotations and emerging trends in biomedical literature; Las Vegas, Nevada, USA: ACM. pp 954–962.
23. Alako B, Veldhoven A, van Baal S, Jelier R, Verhoeven S, et al. (2005) CoPub Mapper: mining MEDLINE based on search term co-publication. BMC Bioinformatics 6: 51.
24. Frijters R, van Vugt M, Smeets R, van Schaik R, de Vlieg J, et al. (2010) Literature Mining for the Discovery of Hidden Connections between Drugs, Genes and Diseases. PLoS Comput Biol 6: e1000943.
25. Bizer CCR (2006) D2R Server - Publishing Relational Databases on the Semantic Web. the 5th International Semantic Web Conference. Athens, GA, USA.
26. Anyanwu K, Sheth A (2002) The ρ Operator: Discovering and Ranking on the Semantic Web. SIGMOD Record 31: 42–47.